\documentclass[aps,nofootinbib,groupedaddress]{revtex4}

\usepackage{url}
\usepackage{graphicx}
\usepackage{amsmath,amssymb,amstext,amssymb,amsfonts,amsthm}
\usepackage{hyperref}
\usepackage{appendix}
\usepackage[margin=1.0in,papersize={8.5in,11in}]{geometry}
\usepackage[utf8]{inputenc}
\usepackage{soul}

\usepackage{color}

\begin{document}

\title{The Limited Accuracy of Linearized Gravity}

\author{John T. Giblin, Jr${}^{1,2}$}
\email[]{giblinj@kenyon.edu}
\author{James B. Mertens${}^{3,4,5}$}
\email[]{mertens@yorku.ca}
\author{Glenn D. Starkman${}^{2}$}
\email[]{glenn.starkman@case.edu}
\author{Chi Tian${}^{2}$}
\email[]{cxt282@case.edu}

\affiliation{${}^1$Department of Physics, Kenyon College, 201 N College Rd, Gambier, OH 43022}
\affiliation{${}^2$CERCA/ISO, Department of Physics, Case Western Reserve University, 10900 Euclid Avenue, Cleveland, OH 44106}
\affiliation{${}^3$Department of Physics and Astronomy, York University, Toronto, Ontario, M3J 1P3, Canada}
\affiliation{${}^4$Perimeter Institute for Theoretical Physics, Waterloo, Ontario N2L 2Y5, Canada}
\affiliation{${}^5$Canadian Institute for Theoretical Astrophysics, University of Toronto, Toronto, ON M5H 3H8 Canada}

\begin{abstract}

Standard cosmological models rely on an approximate treatment of gravity,
utilizing solutions of the linearized Einstein equations as well as
physical approximations. In an era of precision cosmology, we should ask:
are these approximate predictions sufficiently accurate for comparison to
observations, and can we draw meaningful conclusions about properties
of our Universe from them? In this work we examine the accuracy of linearized
gravity in the presence of collisionless matter and a cosmological constant
utilizing fully general relativistic simulations. We observe the
gauge-dependence of corrections to linear theory, and note the amplitude of
these corrections. For perturbations whose amplitudes are in line with
expectations from the standard $\Lambda$CDM model, we find that the full, general
relativistic metric is well-described by linear theory in Newtonian and harmonic
gauges, while the metric in comoving-synchronous gauge is not. For the largest
observed structures in our Universe, our results suggest that corrections to linear
gravitational theory can reach or surpass the percent-level.

\end{abstract}

\maketitle

\tableofcontents

\section{Introduction}

It has been demonstrated that our Universe is mostly well-described on
large scales by a standard cosmological model, consisting of matter with
properties similar to that of a pressureless perfect fluid
and dark energy with properties similar to that of a cosmological
constant ($\Lambda$CDM). At the same time, structures
are formed through nonlinear interactions on scales only somewhat
smaller than the the Hubble scale--the Universe is decidedly
inhomogeneous. These nonlinear interactions are commonly
modeled using Newtonian N-body simulations, which account for
nonlinearities in the matter sector, while large-scale inhomogeneities
are commonly modeled using linear cosmological perturbation theory.

Such approximate treatments of gravitational interactions help improve
the tractability of calculations, and offer us physical insight into the
dominant gravitational effects. However, the connection between such
approximate treatments and fully relativistic treatments is not often
made nor fully quantified. Yet it is necessary in
order to meaningfully interpret and understand theoretical predictions,
especially given a goal of testing general relativity and alternatives
to it in a cosmological setting.
One concern is that, upon coarse-graining a spacetime, we lose insight
into the underlying properties of the spacetime. For example, this
idea is at the core of the Ricci-Weyl problem \cite{1981GReGr}, which has
been shown to have implications for how observables are interpreted in
an inhomogeneous universe \cite{1706.09383}.
Studies of nonlinear gravitational effects have also shown that energy
from small-scale gravitational interactions can
have a considerable impact on the cosmological properties of a spacetime
\cite{1203.6478,1412.3865}. Even within a standard perturbative
cosmological framework, it has been shown that neglected
relativistic effects can lead to percent-level or larger corrections 
when computing observables \cite{1708.00492,1610.03351,1612.03726},
providing a means by which we can study
general relativity beyond its dominant behavior.

Numerical relativity offers us a unified framework in which to examine such
questions, providing infrastructure largely agnostic to gauge and matter
content, which, by construction, will exactly account for all gravitational
effects. While the magnitude of effects on both observables and the spacetime
metric have been examined in a standard cosmological setting using a fully
relativistic approach
\cite{1711.06681,1707.06640,1511.05124,1608.04403,1511.01105,1807.01711,1807.01714},
a more systematic examination of the order of magnitude of corrections
to standard linear calculations has not yet been made. To this end,
we compare linear theory to fully general relativistic simulations performed
without reliance on a background model or any perturbative assumptions.
We evolve a collisionless stress-energy source and cosmological constant, with
energy densities corresponding to values inferred from observations.
We then explicitly demonstrate the gauge dependence of the magnitude of
corrections to linear theory, and note the approximate order of magnitude
in several commonly used gauges as a function of scale.
We examine both overdensity amplitudes predicted by the $\Lambda$CDM model,
as well as larger density contrasts consistent with observed structures.

We then examine in detail two measures of the inaccuracy of linear theory: we
first compute the level of violation of the linearized Einstein field
equations, and second compute the magnitude of disagreement of background
FLRW quantities with spatially averaged quantities.
The first of these measures provides us with a way to test the accuracy of
linear gravitational theory. In linear theory, fields are often transformed
between gauges using linearized definitions of gauge transformations, thus
a solution obtained in one gauge can be mapped to another gauge. Precise field
values are then expected to differ between gauges, but in any gauge,
corrections to linear theory are expected to be of order
$\mathcal{O}(h^{2})$ for a metric perturbation amplitude $h$.
On the other hand, the average behavior of spacetimes is generally expected
to agree with an FLRW model, thus we directly compare the average properties
of our simulations to an FLRW model. We observe the gauge-dependence of this,
and quantify the difference we observe across a
range of physical scales.

When simulating density perturbations of statistically common amplitudes, we
find that in comoving synchronous gauge, the difference between fully relativistic
and linear models of overdensities can be much larger than
Newtonian and harmonic gauges, where the difference can be extremely
small. For some of the largest structures in our Universe that have been
observationally well-characterized, we find that nonlinear effects can approach
the percent level even in harmonic and Newtonian gauges, while linear theory
is entirely unable describe the synchronous gauge metric. These findings coincide
with expectations: it is well-known that the amplitude of metric perturbations can
vary significantly between gauges or ``slicing conditions'' \cite{Bardeen:1980kt},
as noted in other approximate and analytic treatments \cite{1407.8084,1706.09309}
as well. For density perturbations of cosmologically common amplitudes in synchronous
gauge, the metric amplitude scales roughly with the density contrast $\delta_\rho$,
and can therefore become quite large, while in harmonic slicing and a
quasi-Newtonian gauge, metric perturbations remain small. The amplitude of metric
perturbations in the presence of extreme structures, on the other hand, can be
considerably larger than expected.

We begin in Section~\ref{sec:methods} by discussing the various methods
we employ in order to obtain our results.
In Section~\ref{sec:metric} we briefly discuss the numerical relativity
formulation we use to evolve Einstein's equations. In Section~\ref{sec:matter}
we discuss the formalism we use to evolve collisionless matter
and some numerical details. In Section~\ref{sec:ICs} we describe the 
initial conditions we use in order to maintain consistency with
standard cosmological calculations, and in Section~\ref{sec:gauges} we
provide some discussion of the different gauges we use. We conclude
by detailing our results in Section~\ref{sec:results}, first describing
the behavior of nonlinear corrections for various ``mode-in-a-box''
simulations with matter overdensity amplitudes predicted by standard
cosmological perturbation theory, and finally performing asymmetric,
fully 3-dimensional runs with observationally-motivated physical
parameters comparable to large voids and overdensities.

\section{Methods}
\label{sec:methods}

Standard cosmology and numerical relativity employ similar, and sometimes
coincident, formulations in which to study the behavior of Einstein's equations.
Common to both is the 3+1 language, in which the evolution of the spacetime metric is
posed as a Cauchy problem, with spatial hypersurfaces evolved forward in time.
We will generally remain within this framework, although will connect to both standard
cosmological perturbation theory, which deals with the linearized Einstein
equations, and formulations of numerical relativity, which are closely related
to the 3+1 decomposition.

\subsection{Metric Evolution}
\label{sec:metric}

We evolve the full Einstein field equations using the BSSNOK formulation
of numerical relativity \cite{Nakamura:1987zz,Shibata:1995we,Baumgarte:1998te}.
This formulation permits use of an arbitrary gauge and stress-energy source,
allowing us to investigate the accuracy of linear theory in a cosmological setting
for an arbitrary gauge, or slicing condition. We begin by
writing two decompositions of the spacetime metric, the 3+1/ADM form, and a
linearized scalar-vector-tensor (SVT) decomposition,
\begin{align}
g_{\mu\nu} & =\left(\begin{array}{cc}
-\alpha^{2}+\beta_{l}\beta^{l} & \beta_{i}\\
\beta_{j} & \gamma_{ij}
\end{array}\right) & {\rm (3+1)}\nonumber \\
 & =\left(\begin{array}{cc}
-1\\
 & a^{2}\delta_{ij}
\end{array}\right)+\left(\begin{array}{cc}
-E & a\partial_{i}F\\
a\partial_{j}F & a^{2}\left(A\delta_{ij}+\partial_{i}\partial_{j}B\right)
\end{array}\right) + {\rm vector\,+\,tensor}\,. & {\rm (SVT)}
\end{align}
Here, $\gamma_{ij}$ is the metric of a spatial hypersurface, and
the parameters $\alpha$ and $\beta^{i}$ are the lapse and shift,
respectively. The lapse and shift are considered gauge variables,
and may be freely chosen.
The SVT scalars $E$, $F$, $A$, and $B$ typically describe the dominant
behavior of a spacetime, especially in a linearized gravity setting.
Vector and tensor modes can formally contribute at linear order in a
perturbative expansion, but respectively decay or remain small in commonly used gauges unless
sourced, and are therefore usually neglected in simulations of large
scale structures. We will not study these contributions here. The linearized Einstein
equations are also found to provide independent equations for the scalar,
vector and tensor modes.

Einstein's field equations can be written in terms of the metric,
the extrinsic curvature $K_{ij}$, its trace $K$, and
stress-energy source terms projected onto the spatial hypersurface
as a system of first-order dynamical equations.
Conformally-related
variants of these fields are evolved in the BSSNOK formulation. A conformal
factor related to the determinant of the metric $\gamma$ is also evolved,
$\phi \equiv \ln( \gamma^{1/12} ) $, which we will refer to later. The metric
fields and extrinsic curvature also satisfy the 3+1 Hamiltonian and
Momentum constraint equations,
\begin{align}
\label{eq:HMconst}
\mathcal{H} & =0=R+K^{2}-K_{ij}K^{ij}-16\pi\rho\nonumber \\
\mathcal{M}^{i} & =0=D_{j}\left(K^{ij}-\gamma^{ij}K\right)-8\pi S^{i}\,,
\end{align}
where $D_i$ and $R$ are respectively the covariant derivative and Ricci
scalar associated with the 3-metric $\gamma_{ij}$, and $\rho$ and $S^i$
are 3+1 source terms, which are given by projections of the stress-energy
tensor $T^{\mu\nu}$ onto spatial hypersurfaces,
\begin{equation}
\rho = n_\mu n_\nu T^{\mu\nu},\,\,\, S_{i} = - \gamma_{ik} n_{\mu} = T^{k\nu},\,\,\, S_{ij} = \gamma_{ik} \gamma_{jl} T^{kl}
\end{equation}
with $n_\mu = (-\alpha, \vec{0})$.

For details
on the code we use, see \cite{Mertens:2015ttp,Giblin:2017juu}, or see
\cite{Baumgarte:2010ndz,Alcubierre:1138167} for a pedagogical introduction
to numerical relativity, including further details on the formulation we use.
Importantly, in the BSSNOK formulation as with numerical relativity
formulations in general, some gauges will be more numerically stable
than others. Slicing conditions commonly found in cosmology tend to
be ill-adapted for numerical evolution, including comoving
synchronous gauge, and relativistic generalizations of Newtonian gauge
(see Appendix~\ref{apx:GNG}).

To alleviate these problems, we sometimes use Z4c constraint
damping while working in these gauges \cite{0912.2920, 1111.2177}. The Z4c
formulation provides a prescription for modifying the BSSNOK formulation in
order to tend the dynamical evolution of the metric towards obeying the non-dynamical
3+1 constraint equations. This does not guarantee better numerical convergence
than the standard BSSNOK formulation, however, in some gauges it is able to suppress the growth
of numerical error  that would otherwise prohibit simulations
from evolving stably at all. This formulation also provides
an alternative to that considered in \cite{1611.07906}, in which the
conformal metric factor was explicitly modified so that the constraint
equations would remain satisfied.

\subsection{Matter Evolution}
\label{sec:matter}

In order to model dark matter, we primarily integrate the Einstein-Vlasov
equations, modeled using an N-body system in harmonic and Newtonian
gauges, and a perfect fluid in comoving-synchronous gauge. The Einstein-Vlasov
equations describe a covariantly conserved phase-space density $f$ along a
trajectory described by an affine parameter $\lambda$,
\begin{equation}
\frac{D f}{d\lambda} = 0\,.
\end{equation}
Simulations developed to solve this equation using N-body techniques within a general
relativistic framework date back to seminal work in 1985 by Shapiro and Teukolsky
\cite{1985ApJ29834S}, who examined the collapse
of structures in a dimensionally reduced setting.
Fully 3+1 N-body simulations appeared as early as 1999 \cite{Shibata:1999va,gr-qc/9905058},
and have recently been employed in, eg., studies of collapse
\cite{1611.07906,1807.11562}. For our N-body simulations, we employ a method similar
to \cite{1611.07906}, making use of additional techniques in order
to accelerate the numerical convergence of simulations. Additional
schemes for modeling collisionless matter have been considered,
\cite{1611.05447,1610.05198,1801.01083,1711.06681,Baumgarte:2010ndz}; 
however a full review of these methods is beyond the scope of our
work.

In comoving synchronous gauge, we find poor convergence of simulations when
using an N-body system. In this gauge, the particles do not move, so small
errors sourced when we compute the density field from these particles can
accumulate over time in a secular manner. This results in an unacceptably
large amount of numerical error, preventing us from obtaining reliable
results. We therefore use a perfect, pressureless fluid in
synchronous gauge, which provides solutions equivalent in the continuum
limit to an N-body simulation when no stream-crossings are present.

The only variable that needs to be evolved for a pressureless fluid in
comoving synchronous gauge is the density. In particular, we can write
a conservation equation for a conformally-related density field $\tilde{D}$,
\begin{equation}
\partial_t \tilde{D} = \partial_{t}(\gamma^{1/2}\rho_0) = 0\,,
\end{equation}
provided a rest-density $\rho_0$ and metric determinant $\gamma$.
We provide further details on evolving this system in
\cite{Mertens:2015ttp}, and focus on describing N-body integration
in the remainder of this section.

Rather than providing a fundamental description of particulate dark
matter, N-body methods can be thought of as a way of discretizing a
phase-space distribution in a manner suitable for numerical
integration. This discrete representation becomes a solution
to the Einstein-Vlasov equations in the
continuum limit, as the spatial resolution and number of particles tend
towards infinity, and particle deposition radius tends towards zero.
We assume that the Einstein-Vlasov system provides a valid
description of cold dark matter on a wide range of
scales, down to scales of order inter-particle spacing for particulate
dark matter. On smaller scales, or in scenarios where the
density distribution is better described by point-like matter sources,
we may hope to gain fundamental insights into relativistic effects
using, eg., black hole lattice studies \cite{1801.01083}.

In a cosmological setting, it is common to start with a uniform density
distribution in phase space that is ``cold'' (3-dimensional).
We therefore begin by examining properties of a phase-space ``sheet'', a density
distribution described by a 3-dimensional sub-manifold of 6-dimensional
phase space. Locations on the sheet can be parametrized by coordinate
labels, or a 3-vector $\vec{s}$. The coordinate location and velocity
of each phase-space point in configuration space will be given by
the functions $\vec{x}(\vec{s})$ and $\vec{u}(\vec{s})$, and the
phase-space density $f(\vec{s})=\mathrm{const}\equiv\rho_{s}$
will be conserved for non-interacting,
collisionless matter.

In order to determine how the fields $\vec{x}$ and $\vec{u}$ evolve
at each point on the sheet, we can integrate the geodesic equations.
In Newtonian gravity, these only require knowing the Newtonian potential,
$\Phi_{N}(\vec{x})$, which in turn is determined by the metric-space
density, $\rho(\vec{x})$, through Poisson's equation. In the relativistic
case, the full metric is required, which is sourced by the 3+1 stress-energy
source terms. The problem is then to determine, given a phase-space
distribution $f$, what the physical density is. The density is
given by the determinant of the Jacobian of the transformation between the
two coordinate systems,
\begin{equation}
\label{eq:jacob_trans}
\rho(\vec{x})=f(\vec{s})/\det\left|\frac{\partial\vec{x}}{\partial\vec{s}}\right|\,.
\end{equation}
Note that $\vec{x}(\vec{s})$ is not always invertible, or the Jacobian
may be infinite at some points; the inverse function $\vec{s}(\vec{x})$
may also be multi-valued, although $\vec{x}(\vec{s})$ should not
be. For further discussion of this, we refer the reader to
Appendix~\ref{apx:caustics} and references therein.

Given this and a coordinate
mapping $\vec{x}(\vec{s})$, we can determine $\rho(\vec{x})$. The standard
Newtonian N-body prescription then discretizes the phase-space distribution
at regular intervals of $\vec{s}$ into ``particles'' with a mass
$m=\rho_{s}ds^{3}$, and considers the mass to be localized near a given
metric-space coordinate $\vec{x}(\vec{s})$. The mass is assigned
to a gridded metric-space density field using one of several mass
deposition schemes. In standard N-body codes,
perhaps the most common deposition scheme is the cloud-in-cell (CIC)
scheme, which assigns particle masses to a weighted average of nearby
metric-space density grid cells, so that $\rho(\vec{x})$ is locally
incremented by some amount $m/dx^{3}$ for each particle. We additionally
implement a tri-cubic spline (TCS) deposition scheme as described
in \cite{1611.07906}. Regardless of deposition scheme, in the limit of
an infinite number of particles with infinitesimal mass, infinite
resolution in metric space, and infinitesimal radius of deposition,
Eq.~\ref{eq:jacob_trans} can be recovered.

General relativistic N-body systems extend this idea to a fully relativistic
setting. For particles $A$ with rest mass $m_{A}$, the source terms
to the 3+1 equations are given by \cite{Baumgarte:2010ndz}
\begin{align}
\rho & =\sum_{A}m_{A}n_{A}W_{A}^{2}\nonumber \\
S_{i} & =\sum_{A}m_{A}n_{A}W_{A}u_{i}^{A}\nonumber \\
S_{ij} & =\sum_{A}m_{A}n_{A}u_{i}^{A}u_{j}^{A}\nonumber \\
S = \gamma^{ij}S_{ij} & =\rho-\sum_{A}m_{A}n_{A}\label{eq:nb-set}
\end{align}
where $W$ is the relativistic Lorentz factor, 
\begin{equation}
W_{A}=\alpha u^{0}=\sqrt{1+\gamma^{ij}u_{i}u_{j}},
\end{equation}
and $n_{A}$ is the relativistic volume element,
\begin{equation}
n_{A}=\frac{1}{W_{A}\gamma(x_{A})^{1/2}dx\,dy\,dz}\,.
\end{equation}
Eq.~\ref{eq:jacob_trans} also picks up a factor of $\sqrt{\gamma}/W$
on the left-hand side of the equation (eg., Eq.~\ref{eq:caustic_soln}).
The geodesic equations can be neatly written in terms of
covariant spatial velocities and contravariant coordinates,
\begin{align}
\frac{du_{i}}{dt} & =-\alpha u^{0}\partial_{i}\alpha+u_{j}\partial_{i}\beta^{j}-u_{j}u_{k}\partial_{i}\gamma^{jk}\nonumber \\
\frac{dx^{i}}{dt} & =\gamma^{ij}\frac{u_{j}}{u^{0}}-\beta^{i}\,,
\end{align}
where $u^{0}$ is given in terms of $u_{i}$ in the expression for
$W$ above. Using this prescription, we can evolve $u_{i}$ and $x^{i}$ at each
point $s^{i}$ according to the geodesic equations, and compute the
density and other stress-energy source fields using Eq.~\ref{eq:nb-set}. In the Newtonian limit where
$u_{i}\ll1$ and $\alpha=1+\Phi_{N}$
with $\Phi_{N}\ll1$, the geodesic equations reduce to the Newtonian
ones, with $\partial_{t}u_{i}=-\partial_{i}\Phi_{N}$ and $\partial_{t}x^{i}=u^{i}$.

One of the drawbacks of N-body methods is the presence of sampling
noise in the stress-energy source fields, which in turn sources
noise in the metric fields. To alleviate this, phase-space element
techniques have been recently been developed \cite{1111.3944,1501.01959}.
One explicit implementation of these methods is to interpolate additional
particles in the phase-space distribution: the fields $x^{i}$ and
$u_{i}$ are known on a grid described by coordinates $s^{i}$, and
can there be interpolated to arbitrary $s^{i}$, providing a way to deposit
additional particles for which the geodesic equations are not explicitly
used to evolve. We take advantage of this, using tricubic interpolation,
depositing additional appropriately weighted masses in order to
increase the smoothness of the density field, and depositing all particles
according to Eq.~\ref{eq:nb-set}.

\subsection{Initial conditions}
\label{sec:ICs}

Synchronous gauge is the only gauge we use that requires a fixed
choice of lapse and shift on the initial surface. The other
gauges we utilize are driver gauges, which impose conditions only
on the time-evolution of the lapse, and can thus be initialized with
the same lapse and shift as synchronous gauge; their subsequent evolution
will be ``driven'' towards a particular slicing condition irrespective
of the initial hypersurface chosen. We therefore choose conditions
with an initial lapse and shift given by the synchronous gauge ones,
\begin{equation}
\text{\ensuremath{\alpha}=1},\,\beta^{i}=0\,
\end{equation}
We then need to solve the Hamiltonian and momentum constraint equations
\eqref{eq:HMconst}.
A choice for the metric itself can be made using the Zel'dovich
approximation \cite{Zeldovich:1969sb}.
In Newtonian gauge, where $\Phi_{N}\equiv E/2$ and $\Psi_{N}\equiv -A/2$, the
Zel'dovich approximation sets the time derivative of the Newtonian
potential to zero, $\dot{\Phi}_{N}=0$. Synchronous gauge potentials
that coincide with this choice in linear theory can be chosen,
\begin{align}
A & =\Phi_{N}\,,\,\dot{A}=B=\dot{B}=0\,,
\end{align}
and from the linearized Einstein's equations in absence of anisotropic
stress we also find
\begin{equation}
\ddot{B}=A\,,
\end{equation}
from which we can also derive the relation $\Phi_{N}=\Psi_{N}$. A choice
of constant trace of the extrinsic curvature, $K={\rm const}$, can then be made
to produce a desired background evolution. Once given this choice of metric, the
density $\rho$ and momentum $S^{i}$ can be solved for using the constraint
equations, and this solution will coincide with the Zel'dovich approximation
in comoving synchronous gauge at linear order, but will now satisfy
the full, nonlinear constraint equations.

We additionally need to determine the primitive stress-energy source variables,
not just the 3+1 sources. Given an arbitrary metric, this can
be accomplished algebraically prior to stream-crossing when the stress-energy
tensor of various components of the universe, including collisionless
matter and a cosmological constant, coincide with that of a perfect
fluid,
\begin{equation}
T^{\mu\nu}=(\rho_{0}+P)u^{\mu}u^{\nu}+Pg^{\mu\nu}\,,
\end{equation}
with equation of state $P=w\rho_{0}$. The 3+1 source terms for this
stress-energy tensor are given by 
\begin{align}
\rho & =\left(\rho_{0}+P\right)W^{2}-P\,,\label{eq:fluid_density}\\
S_{i} & =\gamma_{ij}S^{j}=\left(\rho_{0}+P\right)Wu_{i}\,.\label{fluid_mom}
\end{align}
For multiple species, the stress-energy tensor will be given by the
sum of individual contributions from all species $\mathcal{S}$,
\begin{equation}
T^{\mu\nu}=\sum_{\mathcal{S}}T_{\mathcal{S}}^{\mu\nu}\,,
\end{equation}
and the contribution from each species to the 3+1 sources will be
given by $\rho=\sum_{\mathcal{S}}\rho_{\mathcal{S}}$ and
$S^{i}=\sum_{\mathcal{S}}S_{\mathcal{S}}^{i}$.
Once given a metric, the constraint equations provide a prescription for
$\rho$, however additional information is required in order to specify the contributions
of each species. In the case of a cosmological constant and collisionless
matter, $\mathcal{S}\in\{m,\,\Lambda\}$, we can divide $\rho$ into
two pieces, respectively a constant $\rho_{\Lambda}$, and the remainder
$\rho_{m}=\rho-\rho_{\Lambda}$.

Once $\rho_{\mathcal{S}}$ and $S_{\mathcal{S}}^{i}$ are determined
for each species, the matter sources can be solved algebraically for
the primitive fluid density and velocity fields, $u_{\mathcal{S}}^{i}$
and $\rho_{0,\mathcal{S}}$  For a cosmological constant $\Lambda$,
the equation of state parameter will be $w=-1$, and the solution
is necessarily $\rho_{0,\Lambda}=-P_{\Lambda}=\rho_{\Lambda}={\rm const}$.
For a more general fluid with $w\neq-1$, a solution for $W_{\mathcal{S}}^{2}$
is given by 
\begin{equation}
W_{\mathcal{S}}^{2}=\frac{1-2C_{\mathcal{S}}R_{\mathcal{S}}+\sqrt{1+4R_{\mathcal{S}}C_{\mathcal{S}}\left(R_{\mathcal{S}}-1\right)}}{2\left(1-C_{\mathcal{S}}\right)}\,,
\end{equation}
with
\begin{align}
R_{\mathcal{S}} & =\frac{w}{1+w}\nonumber \\
C_{\mathcal{S}} & =\gamma_{ij}S_{\mathcal{S}}^{i}S_{\mathcal{S}}^{j}/\rho_{\mathcal{S}}^{2}\,.
\end{align}
Eq.~\ref{eq:fluid_density} can be subsequently solved for $\rho_{0,\mathcal{S}}$
and Eq.~\ref{fluid_mom} solved for $u_{i,\mathcal{S}}$. This approach
to setting initial conditions remains valid for an arbitrary equation
of state parameter $w$, including the case of an N-body system in
absence of stream-crossings. For such collisionless matter or a $w=0$
fluid, the solution reduces to $C_{m}=0$, $W_{m}^{2}=1$, and
$\rho_{m}=\rho_{0,m}$, again consistent with linear theory.

The final solution we obtain for collisionless matter and a
cosmological constant is then determined.
Given an initial $\Phi_{N}$ and constants $K$ and
$\rho_{\Lambda}/\rho=\Omega_{\Lambda}$, we have
\begin{align}
\label{eq:icsols}
\gamma_{ij} & = -2\Phi_{N}\delta_{ij} & \rho_{0,\Lambda} & =\Omega_{\Lambda}\rho \nonumber \\
K_{ij}^{\rm TF} & =u_{i,m}=0  & \rho_{0,m} & =\left(1-\Omega_{\Lambda}\right)\rho\,,
\end{align}
where $16\pi\rho = R + 2 K^{2} / 3$.
In order to determine $\rho$ in practice, we specify the BSSNOK conformal
variable $\phi$ as the solution to $e^{4\phi} \equiv \gamma^{1/3} = 1+A=1-2\Phi_{N}$.
Once we have the metric and density field $\rho_{0,m}$, we can find
a Jacobian transformation that satisfies Eq.~\ref{eq:jacob_trans} in
order to set initial particle positions.

Often in numerical literature, additional physical perturbations
are introduced on small scales as numerical resolution is increased.
This procedure can result in confusion between numerical effects
and new physics, leading to difficulty verifying the numerical
accuracy of results, especially when we are interested in resolving
corrections that can have quite small amplitudes. We therefore opt
to study solutions initially described by a single mode in one
dimension, and three modes in three dimensions, and to verify formal
numerical convergence of our solutions with a fixed physical construction.

Some progress can be made analytically determining particle displacements
for a given density field, allowing us to obtain initial conditions
efficiently, as well as to benchmark the accuracy of a more general solver.
Determining a displacement field is difficult in general, as it requires
the density field be both analytically integrable and invertible.
We can nevertheless write down an analytically integrable density field that
is a solution to the constraint equations, in which case determining
particle displacements is reduced to a root-finding problem. We choose the
conformal factor $\phi$ to be
\begin{equation}
\phi = \log [ 1 + A \sin(2\pi x / L) ]
\end{equation}
and the extrinsic curvature (expansion rate) $K = K{\rm FLRW}$
to determine the density field using Eq.~\ref{eq:icsols}.
The integral of the density field can then be written,
\begin{align}
\int dx \sqrt{\gamma} \rho_{0,m}(x) = & \frac{1}{1920 \pi L^2} \Big[
    120 \pi  x \left(16 \pi A^2+\left(5 \left(A^4+18 A^2+24\right) A^2+16\right) L^2 \bar{\rho}_{m}\right) \nonumber \\
  & + A L \left(A \left(L^2 \bar{\rho}_{m} \left(-5 A^4 \sin \left(\frac{12 \pi  x}{L}\right)-72 A^3 \cos \left(\frac{10 \pi  x}{L}\right)+45 \left(A^2+10\right) A^2 \sin \left(\frac{8 \pi  x}{L}\right) \right.\right.\right. \nonumber \\
  & + \left. \left. 200 \left(3 A^2+8\right) A \cos \left(\frac{6 \pi  x}{L}\right)-225 \left(A^4+16 A^2+16\right) \sin \left(\frac{4 \pi  x}{L}\right)\right)-480 \pi  \sin \left(\frac{4 \pi  x}{L}\right)\right) \nonumber \\
  & \left. -240 \left(3 \left(5 \left(A^2+4\right) A^2+8\right) L^2 \bar{\rho}_{m}+8 \pi \right) \cos \left(\frac{2 \pi  x}{L}\right)\right)
 \Big] \nonumber \\
 \simeq & \bar{\rho}_{m} x - \frac{A \left(3 L^2 \bar{\rho}_{m}+\pi \right)}{\pi  L} \cos \left(\frac{2 \pi  x}{L}\right) + \mathcal{O}(A^2) \,,
\end{align}
where $\bar{\rho}_{m} = (1.0 - \Omega_\Lambda) K^2 / 24 \pi$, and from
which the displacement field is $x(s)$ solved for.

For more generic initial conditions, especially in three spatial dimensions,
we require a method to displace particles to reproduce an arbitrary
density field. In standard N-body simulations, displacements are commonly
computed within perturbation theory. However, such a solution will
not obey the constraint equations exactly, resulting in a small but
measurable amount of violation of the full GR constraint equations.

A more general way of ensuring the constraint equations are fully
satisfied is a diffusion method, in which particles are drifted until a
desired density field is produced. The idea behind this method has been used
in the past in various settings, for example equalization of cartographic
data according to properties such as population density \cite{Gastner7499}.
We begin with an approximate guess of particle positions $x_{g}^{i}(s^{i})$
which generate a density field $\rho_{g}$, and adjust particle positions
in order to send $\rho_{g}\rightarrow\rho_{0,m}$. Phrased in terms
of the difference of these, we wish to equalize $\Delta\rho\equiv\rho_{0,m}-\rho_{g}$,
so that $\Delta\rho\rightarrow0$. We can drift particle positions
along gradients in the field $\Delta\rho$ according to 
\begin{equation}
\dot{x}_{g}^{i}=\eta_D\partial_{i}\left(\Delta\rho\right)\,,
\end{equation}
for a given diffusion strength $\eta_D$. The coefficient $\eta_D$
need only be chosen so that the solution converges; larger values
of $\eta_D$ may result in faster convergence, but can also result
in instabilities if particles ``overshoot'' final positions along
their trajectories.

\subsection{Gauge Choices}
\label{sec:gauges}

In this section we review some common gauge choices, drawing parallels
between gauge choices in approximate treatments and fully relativistic
counterparts in a $3+1$ language.

Comoving synchronous gauge (also referred to as geodesic slicing,
and which refer to more concisely as ``synchronous gauge'' throughout this work,)
is well-defined in both a fully relativistic setting and in linear
theory, ie. can be defined without requiring approximations. This
choice corresponds to 
\begin{equation}
\alpha=1,\,\beta^{i}=0
\end{equation}
in a $3+1$ language, or choosing $E=0$ and $F=0$ in the SVT language.
Some remaining ambiguity exists in how the potentials $A$ and
$B$ are chosen, although this is irrelevant as long as observables
are computed in the end. While we do not compute observables in
this work, this ambiguity remains irrelevant insofar as we are
interested in studying the accuracy of linear theory in describing
relativistic properties of the spacetime in specific
gauges, rather than drawing conclusions about physical quantities
or observables.

Harmonic slicing can be viewed as a foliation that tends to evolve towards
maximal slicing, or a driver condition for maximal slicing, where maximal
slicing is defined by choosing a lapse such that $K=0$ \cite{Smarr:1977uf}.
This gauge choice coincides with the time slicing used in the harmonic
formulation of Einstein's equations, where coordinates themselves satisfy
$\Box x^\mu = 0$. However, we maintain zero shift, so the condition is not
entirely equivalent.
In an FLRW setting, this condition is also modified so that rather
than being driven to zero, the value of $K$ will be driven to that of
a reference FLRW value \cite{1306.1389}. A constant-$K$ condition
is sometimes referred to as a uniform expansion gauge; here we are considering
a driver version which can approximate this condition. The resulting
expression is
\begin{equation}
\partial_{t}\alpha=-\eta_{H}\alpha^{2}\left(K-K_{\rm FLRW}\right)\,,
\end{equation}
where $K_{\rm FLRW}$ is the trace of the extrinsic curvature we wish
to drive towards, and can be freely chosen. We use a coefficient $\eta_{H}=1$
as is common, but also note that as $\eta_{H}\rightarrow\infty$, uniform expansion
should be recovered, reducing to the maximal slicing condition when
$K_{\rm FLRW}=0$ \cite{Baumgarte:2010ndz}. The constant-$K$ limit has also been
studied in detail in a linearized cosmological setting in \cite{Bardeen:1980kt}.

Lastly, we wish to consider a gauge analogous to Newtonian gauge, commonly
found in a linearized, scalar-only context \cite{Mukhanov:1990me}.
This condition can be imposed on the scalar part of the metric by setting
the scalar potential $B=0$, or choosing the scalar part of $K_{ij}$ to be
purely traceful. Unfortunately, although we refer to this gauge as
``Newtonian'' in order to make the analogy explicit, there
is no exact generalization of Newtonian gauge in a fully relativistic setting.
A more general choice is to impose a zero shear (or ``isotropic expansion'')
condition on the spatial metric \cite{1995STIN}, however this condition relies
upon both a choice of background and an assumption of linearity. A relativistic
generalization of this condition would involve imposing this condition
on the anisotropic contributions to the extrinsic curvature, requiring
\begin{equation}
K_{{\rm TF}}^{ij}=0\,,
\end{equation}
however not enough gauge freedom exists to fully enforce this condition.
Instead, we consider a minimal shear driver condition in order
to reproduce the zero-shear condition to some approximation. Our choice
of this minimal shear condition is given by
\begin{equation}
\partial_{t}\alpha = \eta_G \left( \frac{2}{3} \nabla^2 \alpha - \frac{1}{\nabla^2} \sum_{i,j} \partial_{i}\partial_{j} R_{ij}^{{\rm TF}}\right)\,,
\end{equation}
which we derive and explore additional properties of in Appendix~\ref{apx:GNG},
and for which we will extrapolate to the $\eta_G \rightarrow \infty$ limit.
Although we refer to this Newtonian gauge, a more appropriate name is perhaps
a ``minimal-shear driver gauge''.

\section{Results}
\label{sec:results}

We now proceed to examine the accuracy of linear theory in several ways. We
do so by numerically obtaining full, general relativistic solutions, and
examining to what degree the linearized Einstein field equations are satisfied.
In particular, we check the linearized, trace-free spatial part of Einstein's
equations,
\begin{equation}
G_{ij}^{{\rm TF}}-8\pi T_{ij}^{{\rm TF}}\,.
\end{equation}
For a universe containing only collisionless matter and dark energy, the 
scalar contributions to this equation give rise to the
expression \cite{Weinberg:2008zzc}
\begin{equation}
\label{eq:viol:V}
(\partial_i\partial_j)^{\rm TF} \mathcal{V} \equiv (\partial_{i}\partial_{j})^{\rm TF}\left(E+A-a^{2}\ddot{B}-3a\dot{a}\dot{B}+2a\dot{F}+4\dot{a}F\right) \simeq 0\,.
\end{equation}
Involving no matter terms at linear order for cold dark matter and a cosmological
constant, this constraint can be viewed as purely gravitational, expected to
remain valid even in the presence of large density contrasts so long as
the metric potentials themselves are small. Neglected stress-energy contributions
to $\mathcal{V}$ come from anisotropic stress, and are of order
$\mathcal{O}(v^2/\partial^2)$. In
comoving synchronous gauge where $E=F=0$, this expression places a constraint
on the potentials $A$ and $B$, while in a Newtonian setting where
$B=F=0$, this enforces $E=A$, or $\Phi_{N}=\Psi_{N}$, a condition explicitly
enforced in much of cosmological literature. Violation of this constraint
is referred to as gravitational slip, and is often looked for as a
signature of modified gravity models \cite{0802.1068}. However, given an
exact solution to Einstein's equations, the linear constraint equations will
not be perfectly satisfied; it is therefore important to check how well this
expression holds lest an observed violation be mistaken as a sign of failure
of general relativity.

We report the violation, $\mathcal{V}$, relative to the root
sum of the squares of the terms comprising it, which we denote $[\mathcal{V}]$.
There is freedom to choose the zero-mode of $B$ up to an arbitrary time-dependent
function; we choose this to be zero. This quantity will still depend on the
FLRW background chosen, so we note
that we use averaged quantities to construct $a$ and its time-derivatives,
in order to decouple the sensitivity of this expression to
the question of how closely averaged quantities follow an FLRW model.

We can nevertheless explore this last point, sometimes referred to as the
fitting problem or ``backreaction'', by checking to see how well FLRW model
parameters agree with spatially averaged quantities. We explore this question
by looking at the average expansion rate on spatial slices and comparing to
the expected FLRW value,
\begin{equation}
\label{eq:viol:K}
\mathcal{K} \equiv 1 + 3 \frac{ H_{\rm FLRW} }{ \left< K \right> } \simeq 0\,,
\end{equation}
where the average is volume-weighted,
\begin{equation}
\left< K \right> = \frac{\int d^3 x \sqrt{\gamma} K}{\int d^3 x \sqrt{\gamma}}\,.
\end{equation}
This quantity will of course depend on the chosen slicing condition--here,
we are interested in the degree to which $\left<K\right> \simeq -3H_{\rm FLRW}$
in gauges commonly used for interpreting cosmological dynamics. Although we
do not directly compute observables in this work, should an observable coincide
with the choice of gauge (eg. the proper time observable \cite{1305.1299}
coincides with the proper time slicing condition of synchronous gauge),
the results we present can contain implications for observable quantities.

We also compare results to the metric perturbation amplitude $h$ by
computing
\begin{equation}
\label{eq:h_defn}
h \equiv \frac13 ( {\rm perm}|\bar{\gamma}_{ij}| - 1 )\,,
\end{equation}
with $|\bar{\gamma}_{ij}| = |\gamma_{ij}|/{\gamma}^{1/3}$, which is the absolute
value of the conformally related BSSNOK metric, and ${\rm perm}$ the permanent
matrix operator, which we use to reduce the expression to an overall measure of
magnitude.

Literature largely anticipates that cosmological systems are
well-described by linear theory on scales larger than $\mathcal{O}(10)$ Mpc,
albeit with appreciable mode coupling on scales up to $\mathcal{O}(100)$ Mpc
due to nonlinearities, and excepting extreme large-scale structures which
we will examine later. In Newtonian gauge, nonlinearities are attributed
solely to dynamics of the matter sector rather than the gravitational sector,
so $\delta_\rho$ can be large while $h$ remains small.
As our interest is in the gravitational sector, we have chosen to measure
the accuracy of standard linear calculations using Eq.~\ref{eq:viol:V}, which
is expected to hold even in the presence of large, $\mathcal{O}(1)$
density perturbations in Newtonian gauge. In synchronous gauge, the amplitude of
metric perturbations
is not suppressed relative to the density contrast, and we do not
expect suppression of metric perturbations relative to density perturbations
on these scales, so considerable violation may appear.

We will consider scales on which we can obtain a satisfactory
answer in all the gauges we consider. In particular, we examine modes of
wavelength $L \gtrsim 35$ Mpc. On the other hand, the largest observable
scales are of order the current Hubble scale, $H_{0}^{-1} \sim 4.4$ Gpc, so we
will also consider modes up to this scale. We run a suite of planar-symmetric
(``mode in a box'') simulations across this range of scales. We then extend our
results to full, asymmetric 3D simulations with two goals: first, we examine the
applicability of the planar-symmetric results to a more general context by
comparing the amplitudes $\mathcal{V}$ and $\mathcal{K}$ between a three-dimensional
and planar-symmetric setting, and second, we consider modes with
larger-than-typical amplitudes in the full 3D case, motivated by observations of
extremely large structures.

\subsection{Benchmarks in a planar-symmetric setting}
\label{sec:LEEv}

The planar-symmetric simulations we run
begin at a redshift $z\simeq50$, when the energy density of the
Universe is dominated by cold dark matter in the concordance cosmology. We run
all simulations with initial conditions as described in Section~\ref{sec:ICs},
using identical initial conditions in all gauges. We determine the initial
amplitude of density perturbations using CAMB to compute the synchronous
gauge density power spectrum $P_{\delta\delta}(k)$ \cite{Lewis:2002ah}, from which we
compute RMS density perturbation amplitudes smoothed on a length scale $L$,
\begin{equation}
\frac{ \sigma_{\rho, L}^2 }{ \bar{\rho}^2 } = \frac{1}{2\pi^2} \int k^2 e^{-(k L)^2} P_{\delta\delta}(k) dk\,.
\end{equation}
We use CAMB settings including the Halofit nonlinear matter power spectrum,
$H_0 = 67.5$, $\Omega_b h^2 = 0.022$, $\Omega_c h^2=0.122$, $n_s = 0.965$,
and otherwise default settings to compute $\sigma_{\rho, L}$, and choose
mode amplitudes so the RMS density perturbations agree with this value.

The planar-symmetric runs
have a metric grid resolution of $N_x = 64,\,96,\,128$ with $N_y=N_z=1$, and
number of particles $N_p = N_x^2/8$. We generally find 2nd-order convergence
in the case that error due to particle deposition dominates, and 4th-order
convergence for large physical box sizes in which case timestepping error
dominates. Finite differencing is 8th-order, and is not found to be
the dominant source of error for particle runs, but can be for the fluid
simulations in synchronous gauge. Our final results are Richardson extrapolated
using all pairs of resolutions, and numerical confidence intervals inferred
from the remaining disagreement between extrapolated values. We provide further
discussion and details of numerical convergence in Appendix~\ref{apx:num_err}.

\begin{figure}[htb]
  \centering
    \includegraphics[width=1.0\textwidth]{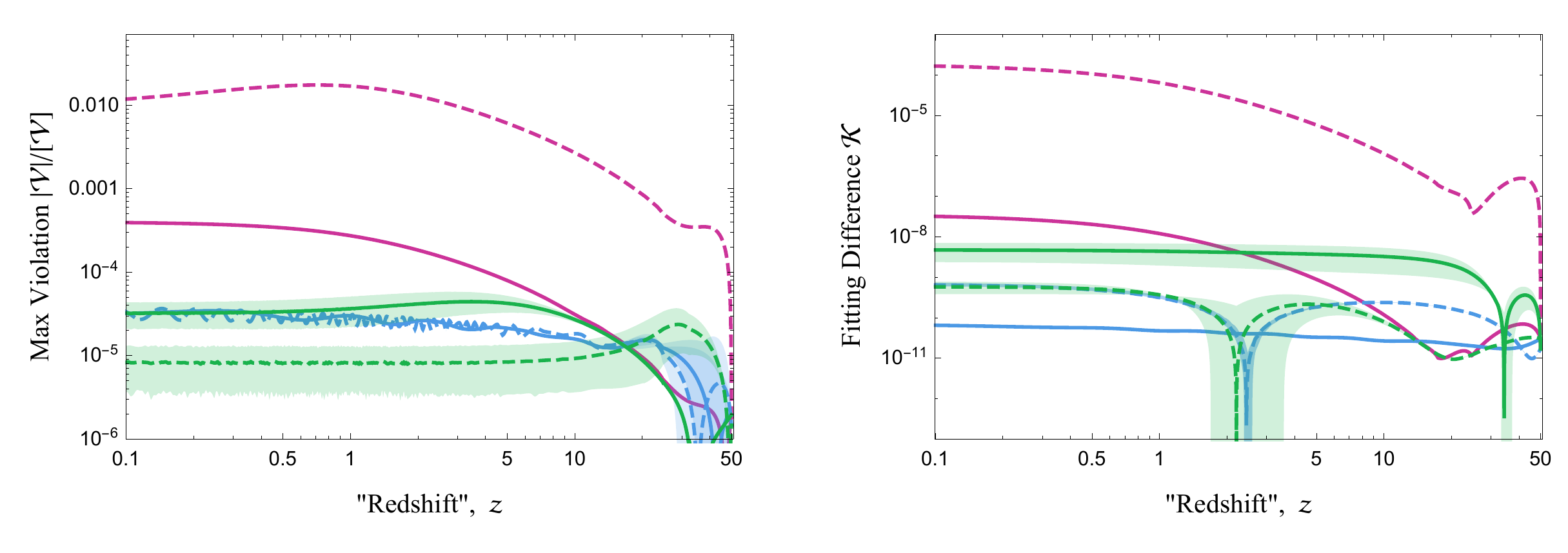}
  \caption{\label{fig:viol:time} Accuracy of approximations for two ``modes'' with concordance cosmology amplitudes in different gauges.
  The left plot shows the evolution of the maximum violation of the linearized Einstein equations, Eq.~\ref{eq:viol:V}, and the right the evolution of the fitting difference, Eq.~\ref{eq:viol:K}. Synchronous gauge results are shown in pink, harmonic in blue, and Newtonian in green. Solid lines indicate results for a $\sim$1 Gpc mode, and dashed for a $\sim$100 Mpc mode. Confidence intervals are indicated for all results using shaded bands, and are not visible when sufficiently small. The Newtonian curves have been extrapolated to the $\eta_G\rightarrow\infty$ limit as described in Appendix~\ref{apx:GNG}. }
\end{figure}

\begin{figure}[htb]
  \centering
    \includegraphics[width=1.0\textwidth]{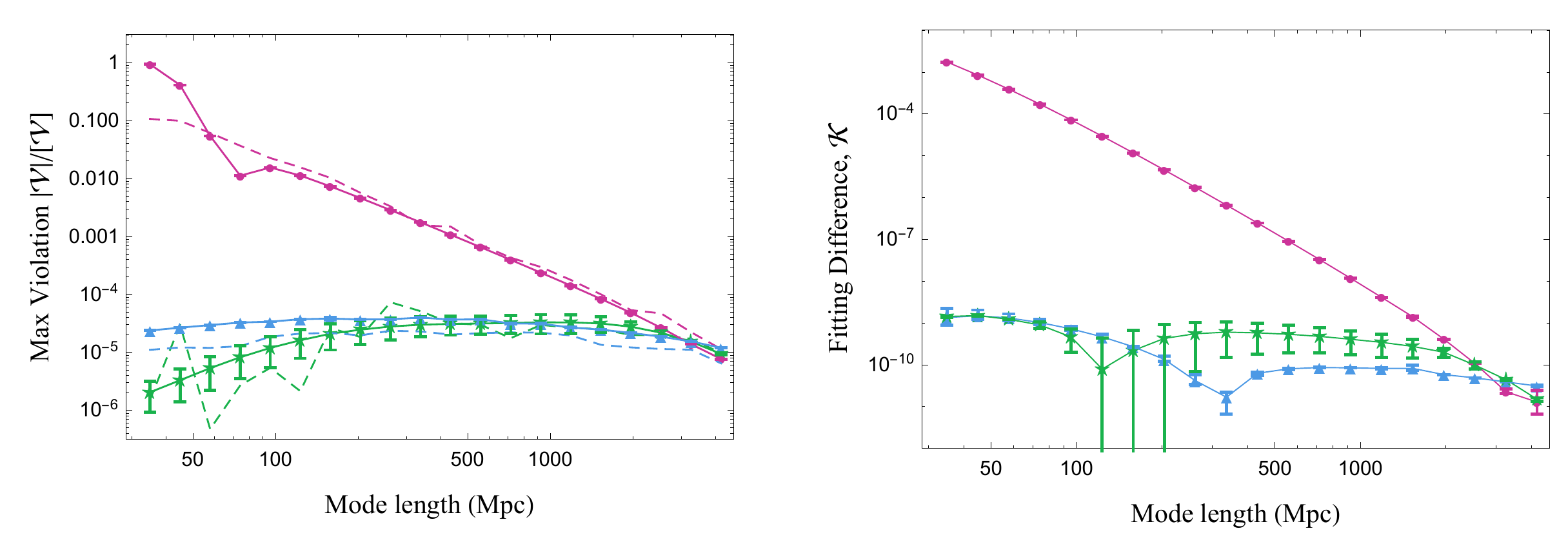}
  \caption{\label{fig:viol:mode} The accuracy of approximations is shown in various gauges as a function of mode wavelength, for concordance cosmology mode amplitudes.  Synchronous gauge results are shown in pink circles, harmonic in blue triangles, and Newtonian (see Appendix~\ref{apx:GNG}) in green stars. Results include numerical confidence intervals and are connected by a solid line, while dashed lines indicate the amplitude $h$ (Eq.~\ref{eq:h_defn}). The left plot shows the difference between a background FLRW Hubble parameter and averaged expansion rate, and the right plot shows the violation of the linearized Einstein equations.  }
\end{figure}

We first examine the degree of violation of the linearized Einstein
equations as described by $\mathcal{V}/[\mathcal{V}]$ in Eq.~\ref{eq:viol:V}.
We show the maximum absolute value of the violation in different gauges
in Figures~\ref{fig:viol:time} and \ref{fig:viol:mode}. The results we find are
consistent with the expectation that
$\mathcal{V} \sim \mathcal{O}(h^2)$ and $[\mathcal{V}] \sim \mathcal{O}(h)$, so
that $\mathcal{V}/[\mathcal{V}] \sim \mathcal{O}(h)$.
Importantly, the degree of violation is found to be scale-dependent and gauge-dependent,
and in fact the violation can be quite large in synchronous gauge where $h$ is
of order $\delta_\rho$, implying the the metric given by linearized calculations
in synchronous gauge will not be accurate when sufficiently small (``nonlinear'')
scales are considered.

We examine both the time-dependence of modes and the scale dependence
of solutions at the end of the runs. The time-dependence is expressed in
terms of the FLRW redshift function of the BSSN conformal factor as
\begin{equation}
z = \exp\left[2\left<\phi_{\rm final}\right> - 2\left<\phi\right>\right] - 1\,.
\end{equation}
Several features can be seen in
both the time-dependence and scale-dependence plots. We interpret these
as transients due to physics in the gauge, initial conditions
which are not purely growing mode solutions, or, especially at early
times, the system transitioning from the synchronous gauge solution to
the preferred slicing in the driver gauges.
Despite the specific dynamics giving rise to these features, we expect that
the order-of-magnitude of these results is fairly insensitive to the precise
initial conditions used.

Turning to examine the overall trends, we indeed find appreciable linearized
constraint violation in synchronous
gauge on scales when nonlinear dynamics begins to become important, or when
$\sigma_\rho/\bar{\rho}$ is no longer small. We generally find
harmonic slicing results to be consistent with small second-order
corrections: violation amplitudes remain below $10^{-4}$ at all
times, in line with expectations from field values $h$. The inexact nature of
Newtonian gauge provides us with a more complicated story, which we provide
detail on in Appendix~\ref{apx:GNG}, however agreement is still found in this
case. As a final note,
there is some expectation that a minimal-shear gauge will behave poorly on
large scales, leading to divergent mode amplitudes. While we have considered
these scales, we have circumvented this difficulty by initializing our
simulations in synchronous gauge, which are then driven
towards other slicings, and may not have sufficient time to appear.
In contrast to the minimal-shear condition, synchronous
gauge is well-behaved on large scales, so there is no such subtlety setting initial
conditions. The $\eta_G\rightarrow\infty$ extrapolation also results in appreciable
error in our Newtonian gauge results, while other numerical errors are considerably
smaller, as seen in other gauges.

We also examine the same data, but check how the degree to which averaged
quantities agree with FLRW quantities. We find
results generally consistent with those seen in the time-evolution:
the behavior of the Synchronous gauge metric is not well-described by
a background FLRW behavior, while the harmonic and Newtonian gauge
metrics are. We may particularly expect that the fitting difference in harmonic
slicing to be small, due to the gauge condition explicitly driving the
solution towards that of an FLRW spacetime. The Newtonian and
synchronous gauge results are also qualitatively consistent with prior
literature looking at backreaction using a partially nonlinear treatment
\cite{1706.09309}, although we do not resolve scales as small as that
work due to the breakdown of synchronous gauge in a fully relativistic
setting, and also expect a fully relativistic calculation to be necessary
in such a nonperturbative setting.

The fitting difference in synchronous and harmonic slicings has also
been examined using fully relativistic simulations
\cite{1511.05124,1807.01714,1706.09309}. In these studies, backreaction
in the sense of violation of the FLRW acceleration equation was been
found to be small, while properties of hypersurfaces such as the average
volume or average expansion rate could show larger deviations from the
average, especially in synchronous gauge.

Because the way we measure the fitting difference is gauge-dependent,
the results we present here, and indeed results found in any gauge,
are only meaningful to either the extent that these quantities
describe observables, or that these quantities are used
as an intermediary step to computing observables. To that end, it is
interesting to consider that Synchronous gauge coincides with a proper
time slicing, making it possible, at least in principle, to pick out
spatial hypersurfaces in this gauge. Newtonian and harmonic slicing
conditions, on the other hand, do not so neatly determine an observable;
no (fully nonlinear) observable properties of a spacetime
coincide with the foliations. Rather, observables are usually constructed
from contributions to the metric and matter fields
whose physical interpretations are gauge dependent.

As a final point, we note that backreaction has also been looked at in
harmonic slicing in the context of black hole lattice simulations. While
the expansion properties of these systems has been found to reproduce
FLRW behavior to some approximation \cite{1307.7673,1306.4055,1404.1435},
it has also been shown that the averaged optical properties of these
spacetimes do not always coincide with expectations from an FLRW
model, even in a homogeneous limit \cite{1611.09275}.
A necessary next step will therefore be examining optical properties of
inhomogeneous spacetimes in a fully relativistic context, especially
in the limit that inhomogeneities are introduced and localized on
increasingly small scales.

\subsection{Benchmarks in a general 3D setting}
\label{sec:3d_results}

As an exploration of over/underdensities with scales similar to the
largest observed structures in our Universe, as well as checking that the
results we find in the planar-symmetric runs are indicative of the order
of magnitude of corrections in a less symmetric setting,
we additionally run fully 3-dimensional simulations in Harmonic and
synchronous gauges. When including perturbations in additional
dimensions we have some expectation that the violation amplitude can
increase: collapse can now occur in these additional directions,
allowing field profiles to further deviate from their original mode-like
profiles than in the 1-dimensional case.

We initialize these simulations at redshift $z=5$ as a superposition
of modes. We follow the prescription described in Eq.~\ref{eq:icsols},
but now using the diffusion method to determine initial particle
displacements. The BSSNOK conformal factor is chosen to be
\begin{equation}
\phi = \sum_i A \sin(2\pi x^i / L)\,,
\end{equation}
where the sum over $i$ is a sum in each direction.

In order to choose relevant and interesting length and overdensity scales, we
consider the case of particularly large voids and overdensities.
For typical overdensity amplitudes on a wide range of scales, there is the expectation
that nonlinear corrections are small in appropriately chosen gauges and a perturbative
approach is therefore well-justified \cite{1610.08882}, something we have explicitly
confirmed above. However, uncommon
structures on large scales with large density contrasts have been observed in
our Universe, for example \cite{1105.3378,1304.2884,1405.1566},
along with even larger but more controversial structures. The density contrast
in these regions can reach tens of percents or more on length scales of up
to a few hundred Mpc. We first use a simulation with parameters comparable to
the void described in \cite{1405.1566}. The diameter of this supervoid is of order
400 Mpc, and density contrast optimistically of order $\delta_\rho \sim -0.2$. We
therefore simulate a box with $L \sim 400$ Mpc and $\sigma_\rho/\bar{\rho} \sim 0.2$
at the end of the simulation. We also run a $L \sim 100$ Mpc simulation with
$\sigma_\rho/\bar{\rho} \sim 1$, comparable to parameters of the Sloan Great Wall
described in \cite{1304.2884}, which has an effective radius of order $50$ Mpc and
reported mass excess $\delta_M$ larger than unity.

\begin{figure}[htb]
  \centering
    \includegraphics[width=1.0\textwidth]{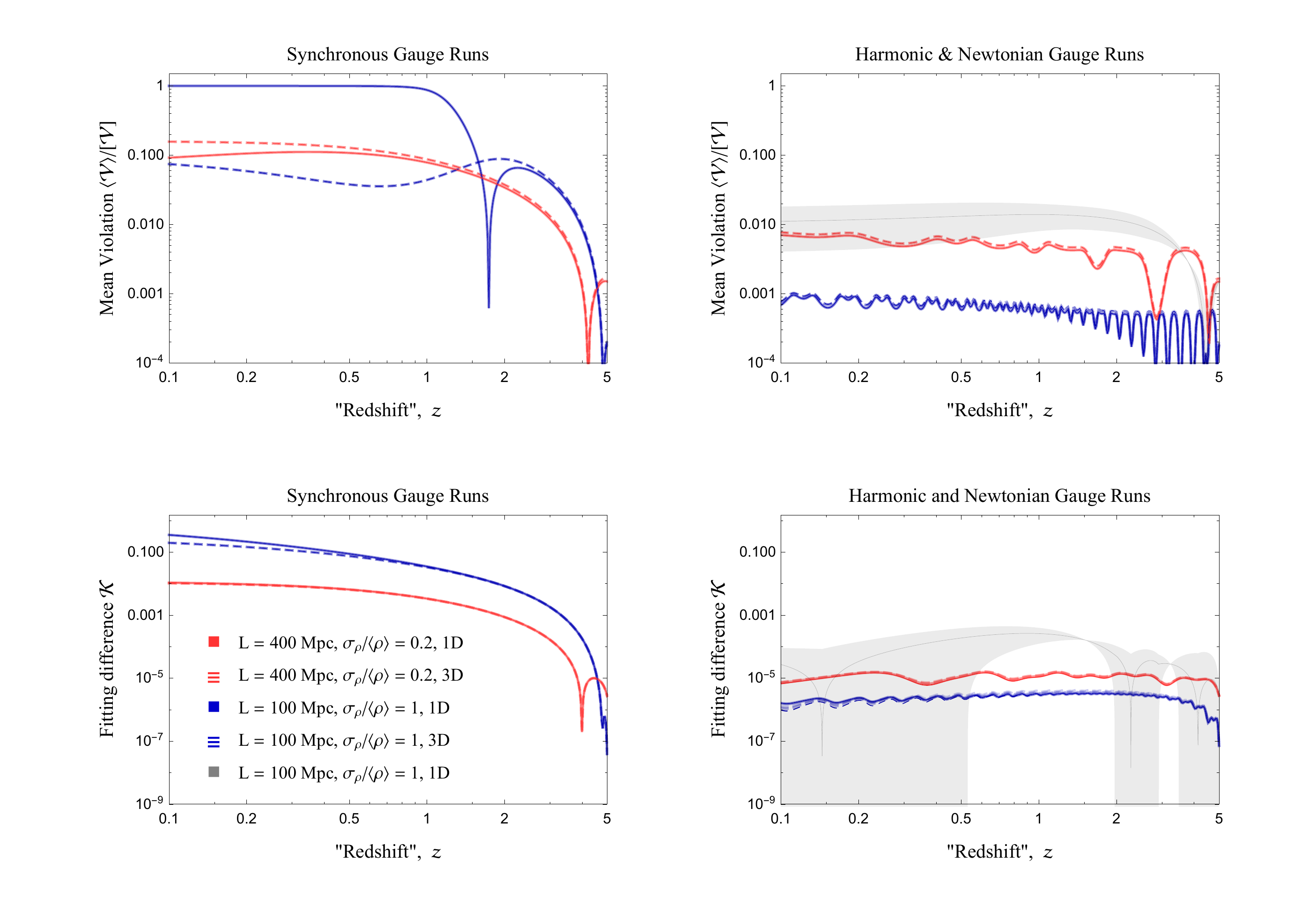}
  \caption{\label{fig:3d_viol} We show the mean violation $\left<\mathcal{V}\right>/[\mathcal{V}]$ and fitting difference $\mathcal{K}$, for planar-symmetric (``1D'', solid) and asymmetric (``3D'', dashed) runs, in synchronous and harmonic gauges as indicated by plot titles, for the large void (red) and overdensity (dark blue), as a function of FLRW redshift. For comparison, the light gray line and band are results and uncertainty from a planar-symmetric simulation using void parameters in Newtonian gauge. Values from individual runs at different resolutions are indicated by thick, light bands, and Richardson extrapolated values thin darker lines. Near-perfect agreement is found in harmonic and synchronous gauges, so individual runs and extrapolated values are nearly indistinguishable. }
\end{figure}

We na\"ively expect the metric perturbation amplitude for these runs to be of order
$\Phi_N \sim 4\pi G L^2 \rho \sim 10^{-3}$. We do not run full, 3D simulations
in Newtonian gauge due to the poor numerical properties of the gauge, but
anticipate given our earlier results and additional 1D tests that fractional
corrections to the Newtonian potential are comparable to harmonic gauge. On
the other hand, we expect the synchronous gauge potentials to be closer to
$\sigma_\rho/\bar{\rho}$.

We run the fully 3-dimensional simulations for the above choices of $L$ and
$\sigma_\rho$, along with planar-symmetric simulations using the same physical
parameters in order to compare results. The synchronous gauge runs utilize
a number of gridpoints $N^3 = 24^3,\, 32^3,\, 40^3$ for all 3-dimensional runs,
and $N = 64, 96, 128$ for all planar-symmetric runs. In harmonic gauge for the
3-dimensional, $L=100$ Mpc runs, we use $N^3 = 16^3,\, 20^3,\, 24^3$ metric grid
points with $N_p = (4N)^3$ particles, and $N^3 = 24^3,\, 28^3,\, 32^3$ with
$N_p = (N^2/4-2N)^3$ particles for the $L=400$ Mpc runs. For the planar
symmetric comparisons, we use $N=24,\,28,\,32$ with $N_p = N^2/2-8N$,
although we also run with higher resolutions as described in Sec.~\ref{sec:LEEv}
in order to examine accuracy and for the Newtonian gauge comparison. We encounter
some difficulty extrapolating Newtonian results to the $\eta_G \rightarrow \infty$
limit for the fitting difference as zero-crossing locations are sensitive to the
driver coefficient, resulting in a large uncertainty in the precise behavior of
$\mathcal{K}$.
We provide further details on the convergence of these runs in
Appendix~\ref{apx:num_err}, and show results from these runs
in Figure~\ref{fig:3d_viol}.

For these runs, we find corrections to linear theory can surpass the expected
$10^{-3}$ level in Harmonic and Newtonian gauge and reach the percent level.
This is still within a perturbative regime, but suggests nonlinear gravitational
effects are beginning to become important for these extreme structures, regardless
of gauge. In synchronous gauge, the corrections become tens of percent up to
order unity, indicating the failure of a perturbative approach to be able to
describe the synchronous gauge metric. We also find that the violation amplitudes
in the 3-dimensional cases are comparable to the planar-symmetric runs, and thus
we expect the planar-symmetric results provide a good indication of the level of
linearized constraint violation.

\section{Conclusions and discussion}

We have examined the magnitude of violations in a standard cosmological
setting using fully relativistic simulations, and found support for
the idea that relativistic corrections scale as $\mathcal{O}(h^2)$ on
large scales. The magnitude of $h$ itself will be gauge-dependent,
will not always be small depending on the gauge in question, and can
be appreciable in any gauge for sufficiently large structures. For codes
that work in synchronous gauge in a linearized context, such as CAMB or
CLASS when run in synchronous gauge, it is therefore inaccurate to directly
infer properties of metric and matter fields in synchronous gauge. However,
because the results of these calculations can be considered as identical
to calculations in a well-behaved gauge under a change of variables,
observables computed from these codes can still be suitably interpreted
as having been performed in a gauge where nonlinear gravitational effects
are small.

It is important to note that we have not directly computed any
observable quantities, but rather examined the accuracy of intermediate
steps used in approximate calculations. The physical interpretation of the
various fields involved will be gauge-dependent in general: for example,
peculiar velocities and matter power spectra will depend on the chosen
slicing. The accuracy with which these effects can be interpreted is
subject to gauge, and as a consequence so are the precise values
of inferred cosmological parameters describing properties of these spacetimes. True observables can still be
constructed by projecting these fields onto past null geodesics.
Relativistic effects involved in this projection are not trivial
to compute, although recent progress has been made both analytically
and numerically \cite{1707.06640,1711.01812}. We can expect that the
accuracy of these projections in a perturbative context will also depend
on gauge: lightcones in comoving synchronous gauge (Lagrangian frame)
can contain quite large distortions,
while these effects will be smaller but still important in, eg.,
Newtonian gauge (Euclidean frame). Observables computed in linear
theory are gauge-invariant only at linear order, and even then linear
gravitational effects are often neglected due to physical or practical considerations
in particular gauges. As a consequence, these observables and corresponding
average properties of the spacetime are perhaps best interpreted as having
been computed under a change of variables in a gauge that maximally
satisfies any approximations made, for example N-body gauge in the case
of Newtonian N-body simulations \cite{1708.07769}.

We are also mindful of the assumptions we have made in this work.
It will be important to work towards relaxing the coarse-graining
operation implicitly performed when utilizing a stress-energy source.
This can be partially addressed by working towards a limit
in which increasingly smaller scales are resolved, and examining any
scaling of nonlinear gravitational effects.
We have also limited the dynamics allowed on large scales by imposing
periodic boundary conditions. For a sufficiently large volume, or
for a volume larger than our observable Universe, the boundary
conditions are less relevant. The challenge then is resolving both
of these scales simultaneously in a fully relativistic setting.

Lastly, we note that due to particle noise and the general expense
of N-body calculations, it can be difficult to obtain reliable
results using N-body simulations. The approach we use here takes
advantage of recent developments in order to improve both performance
and numerical convergence, however cosmological simulations in general
may stand to greatly benefit from the development of methods that
converge at higher order, which can result in a lower
computational cost per accuracy goal.

The agreement we find with linear theory in Newtonian and harmonic
gauges is encouraging for standard cosmological theory in the
context of the late Universe. Yet, we also find that spacetimes
may not always be well-described by linear theory when
sufficiently large overdensities or voids are present in these gauges.
The inability of linear theory to accurately describe the synchronous
gauge metric, on the other hand, implies that a fully relativistic
treatment may generally be necessary to obtain the metric in that gauge.
The large amplitude of nonlinear terms can, for example, explain the
gravitational slip we observed in \cite{1707.06640}; the
amplitude of the slip was $\mathcal{O}(\mathcal{V})$, but it
was computed by evaluating quantities from linear theory using 
the fully relativistic synchronous gauge metric, and was therefore found
to be large. Nevertheless, additional linear-order gravitational
corrections are still often neglected in literature, as in the
approximate expression used in that work.

In this work we have shown the importance of considering linear theory in
the context of a fully relativistic treatment. We have verified in a
fully relativistic setting that nonlinear corrections are of the expected
magnitude for perturbations with amplitudes expected in a concordance
cosmological model in Newtonian and harmonic gauge, and have explicitly
demonstrated that large metric perturbations can be found in the presence
of large density contrasts in synchronous gauge. We have also found that
nonlinear effects can give rise to larger-than-expected metric perturbations
when considering the most extreme structures in our Universe, approaching
percent-level corrections to the metric even in Newtonian and harmonic
gauges. These results suggest care should be taken when attempting to
infer properties of the fully relativistic metric, and in the presence of
large structures where linear theory may begin to fail.

\section{Acknowledgements}
We would like to thank Matt Johnson, Will East and Thomas Baumgarte for discussions
that helped shape this work. This work benefited from the Sexten
Center for Astrophysics workshop on GR effects in cosmological
large-scale structure. This work made use of the High Performance
Computing Resource in the Core Facility for Advanced Research Computing
at Case Western Reserve University, as well as hardware provided by the
National Science Foundation and the Kenyon College Department of Physics at Kenyon College.
This research was supported in part by Perimeter Institute for
Theoretical Physics. Research at Perimeter Institute is supported by
the Government of Canada through the Department of Innovation, Science
and Economic Development Canada and by the Province of Ontario through
the Ministry of Research, Innovation and Science.  JTG is supported
by the National Science Foundation Grant No. PHY-1719652. JBM acknowledges
support as a CITA National Fellow. CT and GDS are supported in part
by a grant from the US Department of Energy.

\bibliographystyle{utcaps}
\bibliography{references}

\begin{appendices}

\section{A toy description of caustics}
\label{apx:caustics}

Formally, the density will become infinite when the Jacobian $\left|\partial\vec{x}/\partial\vec{s}\right|$
is zero, a phenomenon that occurs during shell-crossings. This is
well-known in Newtonian theory \cite{1982GApFD}, but general relativistic
studies of such phenomena are few, typically found in studies of
relativistic collapse \cite{1985ApJ29834S,1611.07906}.

In the presence of an infinite density source, the full, general relativistic
constraint equations imply that curvature scalars should also diverge.
Because this can indicate the presence of a physical singularity, we would
like to obtain solutions to the constraint
equations to gain insight into the behavior of the metric
in the vicinity of a caustic. To this end, we first consider a toy
caustic in a universe with planar symmetry. We choose displacements
\begin{equation}
x(s)=s^{1+p}
\end{equation}
with $p$ a positive integer (setting the dominant term in a series
expansion around a caustic) so that for a constant $\rho_s$,
\begin{equation}
\label{eq:caustic_soln}
\sqrt{\gamma}\rho = \rho_{s}\left|\frac{\partial x}{\partial s}\right|^{-1}=\frac{\rho_{s}}{1+p}\left|x\right|^{-p/(1+p)}\,,
\end{equation}
which diverges when $s=0$, or $x=0$, and the severity of the divergence
is controlled by $p$. Using the constraint equations and choosing an
asymptotically flat and (3-)conformally flat spacetime with $K=0$, or neglecting
cosmological effects on the dynamics, the Hamiltonian constraint equation can be
written in terms of the 3+1 conformal factor $\psi$
(where $\gamma^{1/2}=\psi^{6}$), and obeys
\begin{equation}
{\nabla^{2}}\psi = -2\pi\psi^5 \rho\,.
\end{equation}
In the vicinity of the caustic, we can expand $\psi$ in a power series
for small $x$ and find a solution that scales as
\begin{equation}
\psi \simeq 1 - 2\pi \rho_s \frac{2+p}{1+p} |x|^{\frac{2+p}{1+p}}
\end{equation}
and therefore does not diverge.
The first derivative of $\psi$ will also not diverge, indicating geodesics
can be integrated through caustics without issue, although second
derivatives, and therefore curvature scalars, may diverge. The mass
present within a test region around the caustic also vanishes as
the volume of the test region is decreased, indicating there is infinitesimal
mass at the point of infinite density. Similar behavior may be found
for a zero-dimensional caustic (i.e., radially displaced phase-space
distribution).

For this toy solution, because the divergence of curvature scalars does
not require the metric itself to diverge, the metric can remain in
a weak-field limit. Because it is also possible to integrate geodesics
through these caustics, such solutions do not appear to represent
a singularity in the sense of a black hole.

Although this is a dimensionally reduced example, we do not empirically
find caustics to lead to numerical instabilities or otherwise prevent us
from performing our numerical integration (except in comoving synchronous
gauge / geodesic slicing). The primary drawback we encounter
is a divergent computed Hamiltonian constraint violation due to subtracting
numerically large values. In a standard Newtonian N-body setting, this
difficulty is absent as the linearized constraint equations are directly
used to determine the metric, so the dynamical Einstein equations
are not enforced. In our case where both the metric and
matter are dynamically evolved, we instead check for convergence of field
profiles, and increased localization of the region in which a large GR
constraint violation is observed.

\section{A relativistic generalization of Newtonian gauge}
\label{apx:GNG}

As noted in the text, no true generalization of Newtonian gauge exists
in a fully relativistic setting due to the presence of additional degrees
of freedom. In a planar-symmetric setting as studied here, general
relativity can be directly mapped to Newtonian gravity for a judicious
gauge choice \cite{Mann:1989ik}, however this idea does not generalize
to an asymmetric 3+1 case, and thus we seek a more general condition.
Perhaps the most commonly used generalization of Newtonian gauge is a
``zero-shear'' condition requiring
\begin{equation}
K_{{\rm TF}}^{ij}=0\,,
\end{equation}
although other approaches to map results from Newtonian theory to a
relativistic setting do exist \cite{1708.07769}.
In a linear setting this condition can be reduced to constraints on the
scalar and vector modes. For the scalar modes in the absence of anisotropic
stress, we can compare the Newtonian-gauge line element to
the BSSN one as in the text below Eq.~\ref{eq:icsols}, and combine
with one of the BSSN equations to produce a gauge condition,
\begin{equation}
\dot{\alpha} = \dot{\Phi}_N = -2\dot{ \phi} = \frac{1}{3}\alpha K\,.
\end{equation}
While this resembles the Harmonic gauge condition, it differs by an important
minus sign that results in rapid growth of numerical error.
Due to this growth, we therefore utilize Z4c
constraint damping in order to stably evolve the system.

This choice also will not enforce the zero shear condition beyond linear order.
A generalization suggested by Bardeen \cite{Bardeen:1980kt} is to instead use
a minimal-shear condition, such that
\begin{equation}
D_{i}D_{j}K_{{\rm TF}}^{ij}=0\,.
\end{equation}
This choice reduces to Newtonian gauge in the scalar sector in the linear limit,
and results in a 4th-order differential equation for $\alpha$.
However as solving an elliptic PDE at every timestep is computationally
demanding, and anyways we wish to set initial conditions consistent with
the synchronous gauge ones, we can instead seek a driver condition and look for a
lapse such that the time-evolution equation for $K_{{\rm TF}}^{ij}$ is damped
and not sourced at linear order. Damping of anisotropic stress is already
present to an extent: the equations of motion
\begin{align}
\partial_{t}K_{ij}^{\rm TF}= & \partial_{t}\left(K_{ij}-\frac{1}{3}\gamma_{ij}K\right)\nonumber \\
= & \alpha R_{ij}^{{\rm TF}}-\left(D_{i}D_{j}\alpha\right)^{{\rm TF}}+\frac{1}{3}\alpha KK_{ij}^{\rm TF}-2\alpha K_{ik}^{{\rm TF}}K_{j}^{k{\rm TF}}-8\pi\alpha S_{ij}^{{\rm TF}}\nonumber \\
 & +\beta^{k}\partial_{k}K_{ij}+K_{ik}\partial_{j}\beta^{k}+K_{jk}\partial_{i}\beta^{k}-\frac{1}{3}\gamma_{ij}\beta^{i}D_{i}K-\frac{1}{3}\left(D_{i}\beta_{j}+D_{j}\beta_{i}\right)K\label{eq:ktf_ev}
\end{align}
contain a damping term,
\begin{equation}
\partial_{t}K_{\rm TF}^{ij}\supset\frac{1}{3}\alpha KK_{ij}^{{\rm TF}}\,,
\end{equation}
along with dominant ``source'' terms
\begin{equation}
\partial_{t}K_{\rm TF}^{ij}\supset\alpha R_{ij}^{{\rm TF}}-\left(D_{i}D_{j}\alpha\right)^{{\rm TF}}+{\rm shift}\,,
\end{equation}
where the ``shift'' denotes the contribution from the last line
of Eq.~\ref{eq:ktf_ev}, which we will discuss later.
In order to eliminate the first-order terms,
we can attempt to choose a gauge for which
\begin{equation}
\alpha R_{ij}^{{\rm TF}}-\left(D_{i}D_{j}\alpha\right)^{{\rm TF}}=0\,,
\end{equation}
as any existing $K_{\rm TF}^{ij}$ will naturally decay.
As with the zero-shear condition, this condition
cannot be fully satisfied, but can be minimized by solving
\begin{equation}
D^{i}D^{j}\left(\alpha R_{ij}^{{\rm TF}}-\left(D_{i}D_{j}\alpha\right)^{{\rm TF}}\right)=0\,.\label{eq:min_shear_cond}
\end{equation}
This gauge condition again results in a 4th-order elliptic condition for
$\alpha$. It can be cast into a driver form by using the right-hand side as
an evolution equation for the lapse,
\begin{equation}
\label{eq:full_gr_driver}
\partial_{t}\alpha=\eta D^{i}D^{j}\left(\alpha R_{ij}^{{\rm TF}}-\left(D_{i}D_{j}\alpha\right)^{{\rm TF}}\right)
\end{equation}
for a driver strength coefficient $\eta$. An ordinary Laplacian $D^{2}$
acts as a viscous term, driving field values at a point to the
local average value, resulting in standard diffusive behavior.
The squared Laplacian $D^{2}D^{2}$ can be viewed in a similar way,
but now with an additional frequency-dependent strength, and with
the opposite sign. This minimal shear condition can therefore
be viewed as a diffusion equation, driving the field profile of $\alpha$
towards one that satisfies Eq.~\ref{eq:min_shear_cond}.

Superluminal propagation may exist for this condition, depending on the driver
coefficient and simulation resolution, imposing a strong restriction on the
coefficient amplitude or quite small timestep so that the Courant condition is
satisfied. A future task would be to construct a condition absent of
this requirement. Such a gauge would provide a generalization of Newtonian
gauge that is well-suited to numerical evolution, requiring neither an
especially small timestep nor require solving an elliptic equation
at each timestep. This idea has been explored in a Newtonian setting
\cite{1602.05700}, and the superior scaling properties demonstrated
in that context. Mapping a general relativistic gauge condition to the
hyperbolic equations of motion found in \cite{1602.05700} would provide
a means of interpreting these results in a relativistic context without
requiring a large speed of propagation of gravity.

In order partially to alleviate this problem, we both linearize and act
the inverse Laplacian on the right-hand side of \ref{eq:full_gr_driver} in
order to cast it into a more standard diffusive form, for which $\alpha$
will still be driven towards the desired value,
\begin{equation}
\label{eq:partial_gr_driver}
\partial_{t}\alpha = \eta_G \left( \frac{2}{3} \nabla^2 \alpha - \frac{1}{\nabla^2} \sum_{i,j} \partial_{i}\partial_{j} R_{ij}^{{\rm TF}}\right)\,,
\end{equation}
although we still need to compute an inverse Laplacian. We unsurprisingly find
that this gauge is not stable numerically, but the use of Z4c constraint
damping still allows solutions to be found. However, even after acting
the inverse Laplacian, the timestep required for stability is still quite
small. We nevertheless find this driver condition to perform better than
the choice of Eq.~\ref{eq:full_gr_driver}, or acting the inverse Laplacian
twice (which results in another unstable gauge condition for which the behavior
resembles exponential decay instead of diffusion).

It is also possible to minimize the vector contribution to $K_{ij}$ by
setting the linearized vector contribution to zero. In cosmology,
this choice together with the linearized minimal shear lapse condition comprises
longitudinal or Poisson gauge. A nonlinear generalization of this shift
condition is the minimal distortion shift \cite{Smarr:1977uf}. A conformally
related, somewhat simplified choice for the shift is widely employed
in numerical relativity simulations in a hyperbolic driver form, with
\begin{align}
\partial_{t}\beta^{i} & =kB^{i} \nonumber\\
\partial_{t}B^{i} & =\partial_{t}\bar{\Gamma}^{i}-\eta_B B^{i}\,,
\end{align}
commonly known as the ``Gamma-driver'' condition \cite{gr-qc/0008067,gr-qc/0209102}.
In the limit $k\rightarrow\infty$ and $\eta_B \rightarrow \infty$, and
in a linearized setting, the longitudinal/Poisson gauge condition
should be recovered.

Although the utility of the gamma-driver gauge condition has
been demonstrated throughout numerical relativity literature,
we do not use this shift condition in the simulations
in this work. The gauge choice of $\beta^{i}=0$ that we employ
results in vector modes instead residing in the 3-metric; these modes
should decay on the scales we consider \cite{Weinberg:2008zzc}.

\begin{figure}[htb]
  \centering
    \includegraphics[width=1.0\textwidth]{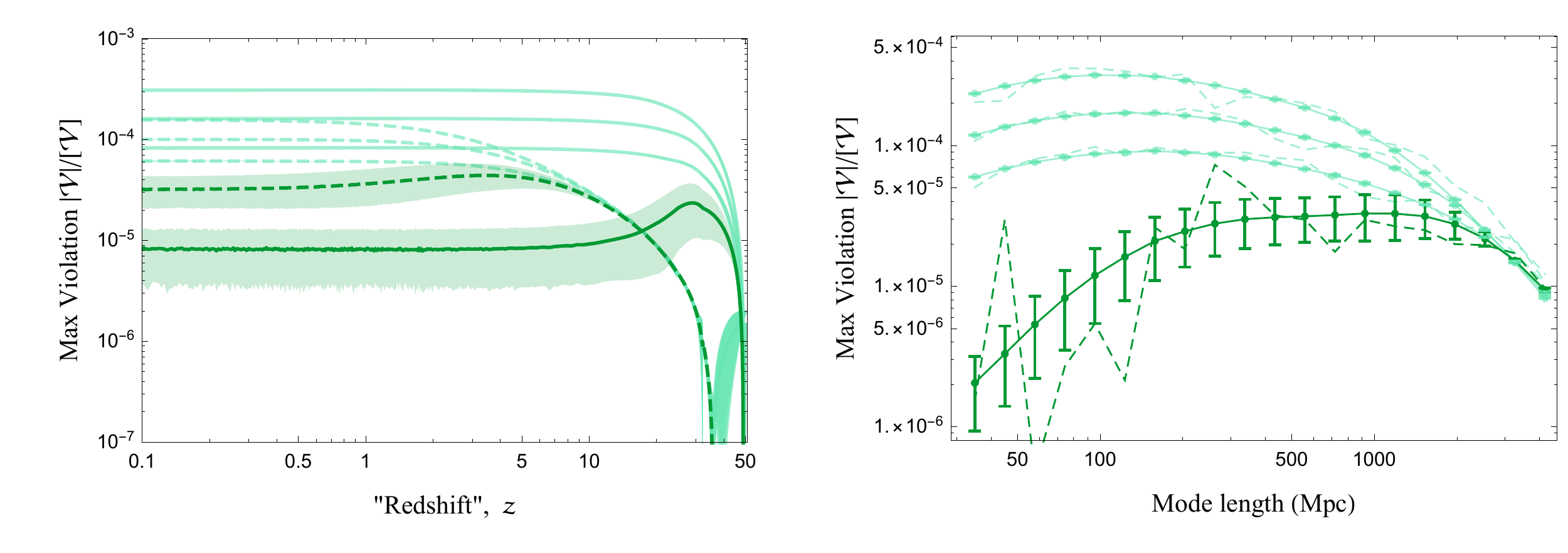}
  \caption{\label{fig:newt_viol} Similar to Figures~\ref{fig:viol:time} and \ref{fig:viol:mode}, we show the relative violation of the linearized Einstein equations, both as a function of mode and FLRW redshift. Light (top three) lines indicate Newtonian results with different values of the driver coefficient ($\eta_G = 0.005,0.01,0.02$ top to bottom), and dark green the extrapolated $\eta_G\rightarrow\infty$ limit as described in the text.
  The left plot shows the violation as a function of redshift, and the right plot shows the violation ``today'' (redshift zero). Other parameters are as in Figures~\ref{fig:viol:time}~and~\ref{fig:viol:mode}. }
\end{figure}

The final results we obtain in the Newtonian driver gauge are sensitive to
the diffusion coefficient in the gauge, with larger coefficients generally
found to result in a smaller amount of linearized constraint violation. In
order to provide a measure
for the linearized constraint violation in the limit that the diffusion coefficient is
infinite, we note that we see the difference between measured linearized violations
follow a power law,
\begin{equation}
\frac{\mathcal{V}}{[\mathcal{V}]} \simeq \left. \frac{\mathcal{V}}{[\mathcal{V}]} \right|_{\eta_G=\infty} + \frac{A}{\eta_G^p}\,,
\end{equation}
with $p \sim 1$ determined empirically. We can then take the limit that
$\eta_G \rightarrow \infty$ in a manner similar to Richardson extrapolation.
Unfortunately this exponent does not work perfectly in all cases: the exponent
inferred from the observed convergence rate varies by tens of percent, leading
to appreciable error bars in the final, extrapolated values. The
extrapolated values nevertheless agree to within the error bars for different
choices of $p$ within this range, so we expect our results including
error bars to robustly quantify the correct order of magnitude of violation of
linear theory in the $\eta_G\rightarrow\infty$ limit.
Results for several choices of $\eta_G$ can be seen in Figure~\ref{fig:newt_viol},
along with the mean violation extrapolated using all pairs of runs,
and error inferred from the standard deviation of this distribution.

\section{Numerical convergence}
\label{apx:num_err}

We briefly describe numerical convergence of our results, focusing
on the results from Section~\ref{sec:3d_results} in Harmonic gauge, which
make use of the particle code and can be run at high resolutions for comparisons.
As noted in the text, we generally find runs converge at second-order in the grid
resolution $\Delta x$. For sufficiently large timesteps, 4th-order error
from RK4 timestepping can dominate: larger physical box sizes allow
larger physical timesteps taken due to the Courant condition. Error from the 8th-order
finite differencing scheme we use was not usually the dominant error source
for the runs we present in this paper, although may in some cases depending
upon choices of numerical parameters. In order for simulations to numerically converge,
the number of particles per smoothing volume--per configuration-space grid cell in our
case--must also tend towards infinity \cite{1410.4222}. Without imposing this
condition when noise dominates the error, we can find non-convergent and inaccurate
field values. Along with the Courant condition, increasing particle number per grid
cell in order to reduce noise poorly impacted scaling in a higher-dimensional setting.

\begin{figure}[htb]
  \centering
    \includegraphics[width=1.0\textwidth]{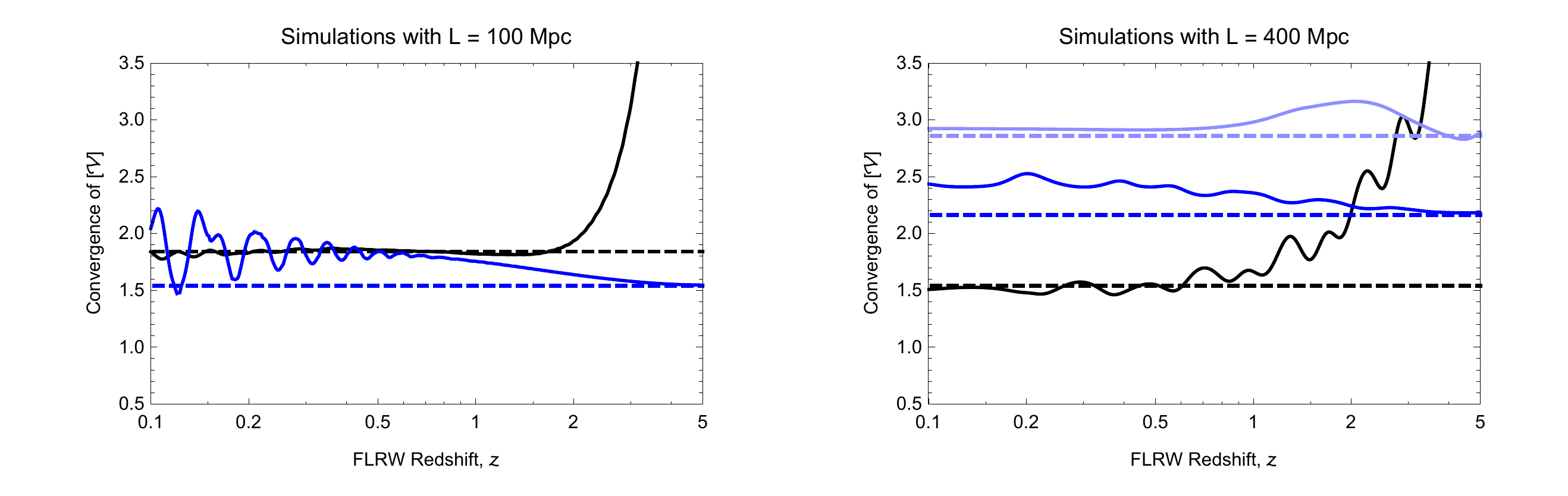}
  \caption{\label{fig:num_err} Numerical convergence of the violation magnitude $[\mathcal{V}]$. Dashed lines indicate theoretical 2nd-order convergence, and solid lines computed convergence. Black lines indicate 3D/asymmetric runs, and dark blue lines ``1D''/planar-symmetric runs. Light blue lines indicate results from planar symmetric Newtonian gauge runs with $\eta_G = 0.02$, while all other runs used harmonic slicing. Resolutions are as described in the text of Section~\ref{sec:3d_results}.
  }
\end{figure}

In Figure~\ref{fig:num_err}, we show the convergence rate of the violation magnitude
$[\mathcal{V}]$ with resolution,
\begin{equation}
\mathcal{C} = \frac{[\mathcal{V}]_{\Delta x_1} - [\mathcal{V}]_{\Delta x_2}}{[\mathcal{V}]_{\Delta x_2} - [\mathcal{V}]_{\Delta x_3}}\,,
\end{equation}
for runs with various resolutions $\Delta x_1 > \Delta x_2 > \Delta x_3$.
Parameters of these runs are as presented in Section~\ref{sec:3d_results}.
In all cases, the convergence rate is found to be second-order in $\Delta x$.
Although not shown explicitly, similar convergence is found for other fields.

For the perfect fluid evolved in synchronous gauge, we find convergence in
agreement with the 8th-order finite difference method used for sufficiently small
timesteps, and 4th-order convergence in the case that error due to
timestepping is dominant. Synchronous gauge results are also Richardson
extrapolated assuming appropriate convergence.

The error bars or ``confidence intervals'' we report come from
the distributions of extrapolated values. We typically run at three
resolutions and can therefore produce three extrapolated values using
three unique pairs or runs, and one extrapolation using all three runs.
The extrapolated values typically agree well, except in the
case of zero-crosings of $\mathcal{V}$ or $\mathcal{K}$. In these cases,
uncertainty is dominated by uncertainty in the precise time of the zero
crossing, which in turn is highly sensitive to truncation error. However,
in these cases, the computed violations were also found to be
uninterestingly physically small.

\end{appendices}

\end{document}